\documentclass[aps,pra,twocolumn]{revtex4}

\usepackage{graphicx,latexsym,color}

\def\be{\begin{equation}}

\def\ee{\end{equation}}

\def\bea{\begin{eqnarray}}

\def\eea{\end{eqnarray}}

\def\bi{\begin{itemize}}

\def\ei{\end{itemize}}

\begin{document}
\hbadness = 10000

\title{Versatile compact atomic source for high resolution dual atom interferometry}

\author{T. M\"{u}ller, T. Wendrich, M. Gilowski, C. Jentsch, E.M. Rasel, and W. Ertmer}
\affiliation{Institut f\"{u}r Quantenoptik, Leibniz Universit\"{a}t Hannover,
Welfengarten 1, D-30167 Hannover, Germany}

\date{\today}

\begin{abstract}
We present a compact $^{87}$Rb atomic source for high precision dual atom interferometers. The source is based on a
double-stage magneto-optical trap (MOT) design, consisting of a
2-dimensional (2D)-MOT for efficient loading of a 3D-MOT. The accumulated atoms are precisely launched in a horizontal moving molasses. Our setup generates a high atomic flux ($>10^{10}\,$atoms/s) with precise and flexibly tunable atomic trajectories as required for high resolution Sagnac atom interferometry. We characterize the
performance of the source with respect to the relevant parameters of the launched atoms, i.e. temperature, absolute velocity and pointing, by utilizing time-of-flight techniques and velocity selective Raman transitions.

\end{abstract}

\maketitle
\par

\section{Introduction }

In the recent past atom interferometry has become a promising technique for high resolution measurements with applications in fundamental physics and for multidisciplinary purposes. Examples are the measurement of the photon recoil for the determination of the fine-structure constant \cite{Ref1}, the measurement of the gravitational constant G \cite{Ref4} as well as gravimeters \cite{Ref2} and gravitygradiometers \cite{Ref3} or atomic gyroscopes \cite{Ref5}. All these experiments benefit from a variety of well developed techniques to coherently manipulate, cool and trap atoms with laser light \cite{met99}.

One of the experimental key elements of atom interferometry is the source of atomic matter waves. In this article we present the design and study of a compact highly stable atomic source for a transportable atomic Sagnac sensor, i.e. a dual Raman-type atom interferometer based on the Mach-Zehnder geometry \cite{Ref6}. We discuss the performance of the source with respect to the requirements imposed by an atomic Sagnac interferometer using free-falling atoms as an ideal inertial reference. Our design is based on a two stage MOT \cite{Raa87}: A 2-dimensional trap \cite{Ref13} creates an atomic beam for efficient loading of a following 3D-MOT. The trapped atoms are released and launched in a moving molasses, similar employed in atomic fountains (see for example Ref.~\cite{Ref8}) to imprint a precise velocity onto the atoms. This setup comprises the advantages of the formerly used source concepts for atomic gyroscopes, which have been thermal beams \cite{Ref5} or loading a 3D-MOT from a background vapor \cite{Ref7}, i.e. a high atomic flux combined with a well defined atomic velocity. Our source is especially compact compared to other state-of-the-art source concepts like Zeeman-slowers \cite{Ref17} or thermal beam devices.

This paper is organized as follows: In Section~\ref{sec:quelle} we discuss the requirements for a source imposed by a Sagnac atom interferometers, which guided us to our source design. The experimental setup of the atomic source is described in Section~\ref{sec:experiment}. In the Sections~\ref{sec:DoppelMOT} and \ref{sec:atomiclaunch}, we characterize the performance of our realized source with respect to the atomic flux and the launch of the atomic ensemble, respectively. Finally, a conclusion and outlook for the rotational measurements is given in Section~\ref{sec:conclusion}.

\begin{figure}[t]

\centering

\includegraphics*[width=8.6cm]{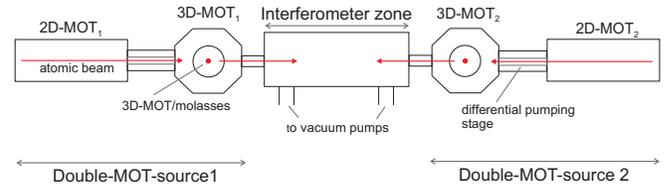}

\caption{Simplified schematic of the experimental setup for the atomic inertial sensor, top view.}

\label{fig:1a}

\end{figure}

\section{Source requirements for atomic Sagnac interferometer}
\label{sec:quelle}
For atom or light interferometers the achievable sensitivity is determined by a scaling factor, i.e. the resulting phase shift induced by the effect to be measured, and the minimal resolvable phase shift of the interferometer. The scaling factor for rotation measurements in atom interferometers can be calculated analogous to the optical Sagnac interferometer. The phase, induced by a rotation $\vec{\Omega}$ of the laboratory system with respect to the interferometer, is proportional to the scalar product with the area $\vec{A}$ enclosed by the interferometer and to the atomic mass $m$ divided by Planks constant $h$~\cite{Ref15}: 

\begin{equation}\label{eq:1}
\Delta\varphi_{rot} = \frac{4\pi m\vec{A}\vec{\Omega}}{h}
\end{equation}

This relation shows, that atomic Sagnac interferometers measure a projection of the enclosed area $\vec{A}$ with respect to the rotation $\vec{\Omega}$. Therefore, the area $\vec{A}$ has to be stable both in absolute value as well as with respect to its relative pointing to the rotation vector to guarantee a well defined scaling factor. For the case of the Raman-type Mach-Zehnder atom interferometer using free-falling atoms~\cite{Ref2,Ref3,Ref4,Ref5,Ref6,Ref7}, that we focus on in this article, the enclosed area is spanned by the atomic drift velocity $\vec{v}_{at}$ and the effective velocity $\vec{v}_{bs}=\hbar\vec{k}_{bs}/m$ transferred by the atom-light interaction during the coherent beam-splitting, redirecting and recombination of the matter waves:

\begin{equation}\label{eq:2}
A = L^{2}\frac{v_{bs}}{v_{at}}sin\vartheta,
\end{equation}

$L$ denotes the spatial separation between the atom-light interaction zones and $\vartheta$ the angle between the atomic drift velocity and the beam-splitting laser used for the coherent manipulation of the atoms. It is notable that the atomic mass drops out of the scaling factor for a given momentum $\hbar\vec{k}_{bs}$.
Apart from the Sagnac effect, the interferometer is also sensitive to accelerations $\vec{a}$ along the direction of the Raman laser beams. Therefore an accelerational phase shift is induced, given by $\Delta\varphi_{acc}=\vec{a}\vec{k}_{bs}T^2$, where $T$ is the time of free evolution of the atomic ensemble in-between the atom-light interactions.

From these relations, a set of requirements can be derived for the design of a compact Sagnac interferometer and for the atomic sources employed in the device. First of all, a high stability and a precise control of the absolute atomic velocity and pointing of the atomic source is required as the atomic velocity enters the scaling factor and determines in combination with the orientation of the beam splitting lasers the orientation of the interferometers sensitive axis.

Moreover, the atomic Sagnac sensor must discriminate between rotations and accelerations. According to equations (\ref{eq:1}) and (\ref{eq:2}) a distinction is possible by a differential measurement based on two simultaneous atom interferometers with reversed atomic beams \cite{Ref5} which results in a change of the sign of the Sagnac phase. In this way, a subtraction or addition of the two interferometers' signals gives access to a purely rotational or accelerational induced phase shift, respectively. In order to apply this method, both sources should emit counter-propagating ensembles of atoms with equal and stable features such as velocity and pointing.

In addition, the ratio of the scaling factors for accelerations and rotations can be modified by changing the atomic velocity $\vec{v}_{at}$. On the one hand, higher atomic velocities and a larger length $L$ of the setup allow to keep a constant enclosed area A while at the same time providing a lower sensitivity for accelerations and inertial noise. As an example, in our interferometer with $L=75\,$mm and $|\vec{v}_{at}|=3\,$m/s the phase shift induced by the earth rotation rate is equal to a phase shift induced by accelerations of $9\times10^{-4}\,$m/s$^2$. Concerning an imperfect suppression of accelerations, this indicates the necessity of isolating the interferometer from the laboratory vibrational noise when aiming for resolutions of fractions of the earth rotation, i.e. $10^{-8}\,$rad/s or better. On the other hand, a compact device with restricted dimensions $L$ on the order of few centimeters recommends the use of cold and slow atoms increasing the ratio of $v_{bs}/v_{at}$ from equation~\ref{eq:2} and thereby enlarging the enclosed area $A$. In this way, an atomic velocity of $1\,-\,20\,$m/s seems to be a proper choice, allowing to maintain the mentioned rotational resolution of $10^{-8}\,$rad/s with a reasonable resolution of the phase shift in the interferometer of about 1mrad.

Another important aspect of the source is the temperature of the atomic ensemble. For inertial measurements the Raman beam-splitters are strongly velocity selective \cite{Ref2,Ref5,Ref7}. As a result, the efficiency of the beam-splitting is reduced for finite temperature. This reduces the effective interferometer signal and leads at the same time to an incoherent background. These effects decrease the sensitivity of the interferometer.

Finally, a high flux of the source, comparable to thermal atomic beams, helps to achieve a good signal-to-noise ratio and enhances the ultimately achievable phase resolution of the interferometer, given by the atomic shot-noise limit. Additionally, a high atomic flux allows the adaption of the preparation time of the atomic sample with respect to the measurement time, to increase the cycling rate of the measurement, minimizing the noise contribution by the Dick effect~\cite{Ref10,Ref11}. Ideally, the best noise characteristic is achieved by performing a continuous measurement~\cite{Thomann}.

\section{Description of the atomic source and experimental setup}
\label{sec:experiment}

For the atomic Sagnac sensor we developed a compact atomic source which consists of a so-called double-MOT-system, where a 3D-MOT is being loaded by a cold and slow atomic beam emitted by a 2D-MOT. After release out of the 3D-MOT, the atoms are launched by a moving molasses towards the interferometer. In this way, the atomic velocity can be controlled with high accuracy by adjusting the frequency difference between the molasses laser beams~\cite{Ref9}. This frequency difference can be referenced to a high quality radio-frequency oscillator in order to achieve a higher precision and stability. Our realization of the moving molasses permits an individual control of the horizontal and the vertical atomic velocity component, respectively. In this way, we can vary the horizontal velocity between 2,5 and $5\,$m/s and, at the same time, compensate for gravity by adjusting the vertical velocity between 0 and $1\,$m/s, resulting in flat parabolic atomic trajectories. The possibility of velocity variation allows more flexible time dependent systematic studies or even the reduction of the impact of time-dependent noise sources. The chosen trajectories and velocity values are a compromise between accelerational noise sensitivity and compactness of the sensor.

In principle, the choice of the flat parabolic trajectories allows the measurement of the all components of the rotation vector $\vec{\Omega}$, like in Ref.~\cite{Ref16}. However, the design is optimized for measurements of the rotation component $\Omega_z$, parallel to gravity.

Additionally, in combination with the 2D-MOT it is also possible to perform a continuous operation of the atom interferometer and benefit from the advantage of this scheme: For this, the second 3D-stage is used only for molasses cooling and redirecting the atoms. For these further manipulations, the use of a 2D-MOT gives good starting conditions, as it is forming a brilliant and already slow atomic beam.

\begin{figure}[t]

\centering

\includegraphics*[width=8.6cm]{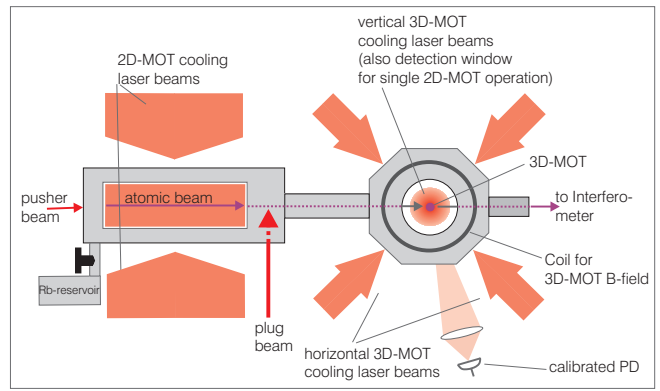}

\caption{Detailed scheme of one double-MOT atomic source, top
view. Not shown are the coils for 2D-MOT magnetic field
generation.}

\label{fig:1b}

\end{figure}

A schematic of the source unit is sketched in Fig.\ref{fig:1b}. One double-MOT consists of two separate chambers
that are connected via a differential pumping stage: The chamber where the 2D-MOT is loaded from Rubidium background vapor and the 3D-MOT chamber at a pressure of about $10^{-8}\,$mbar. The vacuum system
is mainly based on custom made parts to keep the setup as compact as possible. The overall size of one source unit including fiber couplers, beam shaping optics and magnetic coils is about 35$\times$30$\times 25\,$cm$^3$. The non-magnetic chambers are made of aluminum, providing large optical access. The optical substrates are directly sealed to
the individual chambers, either using vacuum compatible glue (for the 2D-MOT), or indium-sealings (for 3D-MOT and
Interferometer). The Rb background pressure in the 2D-MOT is maintained by a heatable vacuum tube containing a Rb
ampule which is connected to the 2D-MOT via a valve. The 2D-MOT and the 3D-MOT chambers have a rectangular and a 8-cornered shape, respectively (in top view). The rectangular shape of the 2D-MOT chamber reflects the transverse cooling volume of 80$\times 18\,$mm$^2$, similar to \cite{Ref12}. The 3D-stage is connected to the interferometer via a vacuum adapter of diameter $30\times 10\,$mm$^{2}$, which can be displaced vertically to allow for even greater variations of the atomic trajectories from almost flat to steep parabolas with a maximum starting angle of about $18^{\circ}$.
The geometry of the 3D-MOT cooling beams is chosen such that the horizontal laser beams are placed at an angle of
45$^{\circ}$ to the 2D-MOT atomic beam axis, whereas the vertical laser beams are perpendicular to it. The magnetic
field for the 3D-MOT is generated with external coils, directly wound onto the chamber (3D-MOT), or in the case for the 2D-MOT, on a
frame rigidly fixed to the chamber, respectively.

Two atomic sources of this kind are integrated into our experimental setup for operating a dual atom interferometer. This setup, shown in Fig.~\ref{fig:1a}, is placed on a single optical breadboard (1.2$\times 0.9\,$m$^{2}$), surrounded by a $\mu$-metal magnetic shield, which suppresses stray magnetic fields by a factor of about 40.

The laser beams for the cooling light are generated by from 2 commercial high power diode laser systems in MOPA configuration (Toptica TA100), delivering $700\,$mW for the 2D-MOT and $500\,$mW for the 3D-MOT. From the 3D-MOT laser four independent laser frequencies are derived for the moving molasses by using acousto-optic modulators (AOMs) in double-pass configuration. Additionally, this laser generates a light beam co-propagating with the cold atomic beam formed by the 2D-MOT, which enhances the forward directed atomic flux thanks to light pressure. This so-called ''pusher''-beam (see Fig.~\ref{fig:1b}) is also frequency controlled using a AOM. A self-made External Cavity Diode Laser (ECDL) produces repumping light for atoms in the lower hyperfine level $|F=1>$.

All lasers are frequency controlled using saturated absorption spectroscopy near the D$_{2}$-line of $^{87}$Rb at $780\,$nm. The laser sources are placed on a separate optical breadboard and the beams are delivered to the main apparatus with polarization maintaining (PM) optical fibers. In all MOT stages the repumping light is superimposed to the cooling light and delivered to the experiment with the same optical fibers, using one individual fiber for each light beam.

The beam shaping optics at the fiber exits, for example beam collimation or polarization retarders, are rigidly fixed to the vacuum chamber. In this way,
the stability of the optical setup is improved with respect to free-space optical paths. Beam diameters are 80$\times 18\,$mm$^2$ for the 2D-MOT, and $30\,$mm for the 3D-MOT.

\section{Single 2D-MOT and combined double-MOT characterization}\label{sec:DoppelMOT}
\subsection{2D-MOT Characterization} \label{subsec:2Dchar} 
In our source concept, the 2D-MOT replaces a Zeeman-slower and serves for generating a large flux of atoms with velocities low enough to be captured by the 3D-MOT or further cooled in the 3D-molasses. The characterization of the 2D-MOT focused therefore on the optimization of the atomic flux and and the velocity distribution of the 2D-MOT's atomic beam by tuning the relevant parameters such as laser power or detuning as well as the magnetic field gradient $\Delta$B.

The flux and velocity distribution were measured by detecting the atomic fluorescence with a calibrated photo diode while the atoms are interacting with a retro-reflected detection laser beam tuned onto the $|F=2>\rightarrow|F'=3>$ transition of the D$_{2}$ line. The detection laser beam of $5\,$mm diameter and $0.5\,$mW power was placed $133\,$mm downstream from the 2D-MOT exit. For the velocity measurement the atomic beam was chopped with a second transverse, running wave laser beam tuned onto the $|F=2>\rightarrow|F'=3>$ transition deflecting the atoms with its light pressure, the so-called ''plug''-beam. Analyzing the time-dependent atom number and integrating over all velocities we can deduce the atomic flux as well as the velocity distribution.

\begin{figure}[t]

\centering

\includegraphics*[width=8.6cm]{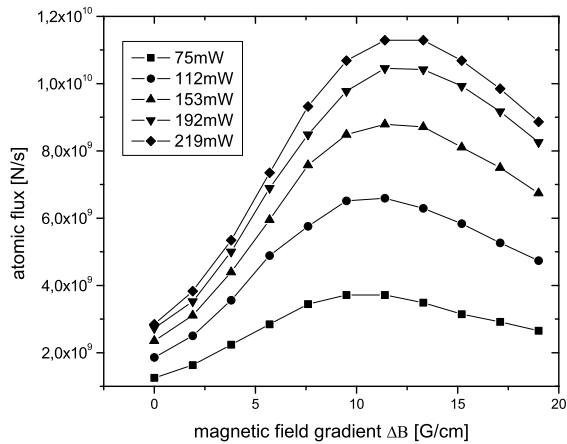}

\caption{2D-MOT atomic flux versus magnetic field gradient for
different total cooling laser power. Other experimental parameters
are: Detuning $\delta=-1.2\,\Gamma$, Rb-reservoir temperature
120$^{\circ}\,$C}

\label{fig:2}

\end{figure}

\begin{figure}[t]

\centering

\includegraphics*[width=8.6cm]{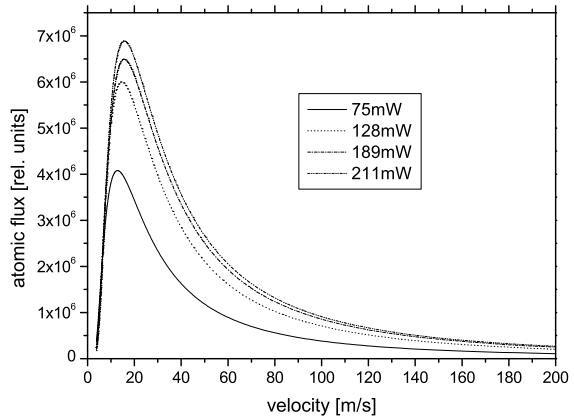}

\caption{Velocity distribution of the atomic beam from the 2D-MOT
for different total laser power. Recorded with
$\delta=-1.2\,\Gamma$, $\Delta$B$=12\,$G/cm, and 120$^{\circ}\,$C
Rb-reservoir temperature}

\label{fig:3}

\end{figure}

The results for a combined measurement is shown in Fig.~\ref{fig:2} for different values of the total cooling laser power and magnetic field gradient. The laser detuning $\delta$ for this measurement was chosen to be $\delta$=-1.2$\Gamma$ (${\Gamma}$: linewidth of the D${_2}$-transition), which was
shown to be optimal in a separate measurement. We infer the optimal magnetic field gradient $\Delta$B for a laser power of $200\,$mW to be about $12\,$G/cm. The measurement shows that the yielded flux can be
enhanced by increasing the laser power and saturates at a value of about $200\,$mW total laser power. The optimal flux achieved in this measurement was $1.1\times 10^{10}\,$at/s. The
corresponding velocity distributions, shown in Fig.~\ref{fig:3}, display that the most probable velocity is less than $20\,$m/s for all laser powers that have been used during the measurement. It shows additionally that increasing the laser power results in higher mean atomic velocities. This is also theoretically predicted by a 2D-MOT model, for example derived in Ref.~\cite{Ref12}.

Previous work by other groups (for example \cite{Ref13}) showed that additional molasses cooling on the atomic beam
axis can also increase the atomic flux and reduce the mean longitudinal velocity. Due to limited optical access on the atomic beam axis we apply a related and simpler scheme with only one single traveling beam. For a detuning $\delta$=+1.8$\Gamma$ to the $|F=2>\rightarrow|F'=3>$ transition and with $0.5\,$mW power this ''pusher''-beam enhances the atomic flux by a factor 4. But as the beam is acting on the longitudinal axis, the mean velocity is also slightly increased.

To estimate the beam divergence, we proceeded as follows: By
cutting a part of the ''plug''-beam with a knife edge transversally,
a part of the atomic beam is not deflected and can still reach the
detection zone. Consecutively moving the edge over the ''plug''-beam, and monitoring the residual detected fluorescence, we can
assign the atomic beam a Gaussian profile. Comparing the measured
atomic beam waists at different distances from the 2D-MOT exit, we
calculate the divergence to be about $30\,$mrad.

The performance of the 2D-MOT comes close to that of thermal sources used for atomic Sagnac interferometry~\cite{Ref5}, while it it allows at the same time to keep the setup compact and transportable.

\subsection{2D-/3D-MOT characterization} \label{subsec:doubleMOT}
In our present interferometer the 2D-MOT serves to quickly load the 3D-MOT, which further compresses and cools the atomic ensemble. Therefore, a major characterization aspect for the combined 2D-/3D-MOT operation was the 3D-MOT loading rate. This loading rate was examined for different parameters (like the rubidium background pressure in the 2D-MOT) and various configurations, for example with and without the ''pushing''-laser beam. We infer the loading rate from the time dependent filling of the 3D-MOT, which is deduced from the atomic fluorescence, by switching the 2D-MOT atomic beam. The switching is performed by mechanical shuttering the light and by tuning the magnetic field of the 2D-MOT. Without the 2D-MOT operating, the influx of atoms into the 3D-MOT is completely stopped and we find no detectable filling of the 3D-MOT by residual background Rb atoms. In the realized setup presented in this paper, the saturation of the absolute 3D-MOT atom number is determined by collisions with background (non Rb) gas atoms with a time constant of about $0.5\,$seconds.

For maximum loading rate, we find as optimal parameters for our 3D-MOT: Detuning $\delta\,=\,-3.5\,\Gamma$ and magnetic field gradient $\Delta$B$\,=\,14.5\,$G/cm. Varying the 3D-MOT laser power, we infer a similar dependence like given in Fig.~\ref{fig:2} for the 2D-MOT. Under usual conditions when serving two complete source units we work with $10\,$mW laser power per 3D-MOT beam, which is at the beginning of saturation.

\begin{figure}[t]

\centering

\includegraphics*[width=8.6cm]{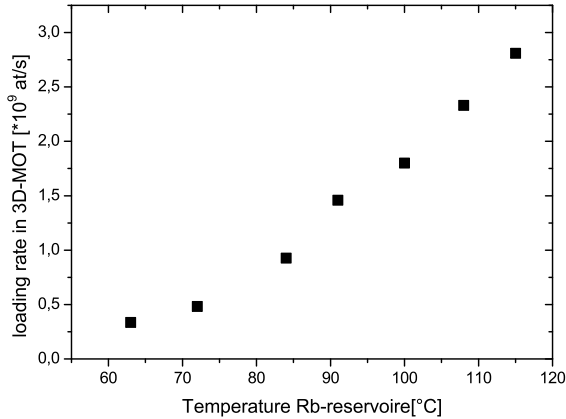}

\caption{Influence of the rubidium vapor pressure on the atomic
flux measured via the 3D-MOT loading rate. Displayed here by the
temperature of the rubidium reservoir, which relates directly to
the pressure. All exp. parameters like mentioned in the text of
section \ref{subsec:doubleMOT}.}

\label{fig:4}

\end{figure}

The dependence of the loading rate on rubidium vapor pressure is displayed in Fig.~\ref{fig:4}. The Rb pressure is adjusted by tuning the Rb-reservoir temperature. The loading rate does not saturate for the chosen temperatures which still leaves room for further boosting the loading of the 3D-MOT (compare for example \cite{Ref13} for a saturation behavior with increasing Rb-vapor pressure).

The 3D-MOT loading rate can also be enhanced by the use of the ''pusher''-beam, see section~\ref{subsec:2Dchar}. For the combined operation of the 2D-/3D-MOT this laser beam has to be aligned with a slight tilt of about 3$^{\circ}$ with respect to the atomic beam, which avoids disturbance of the 3D-MOT. In this configuration, the loading rate can be enhanced by a factor 3.5 with $0.9\,$mW light power in the ''pusher''-beam.

\begin{figure}[t]

\centering

\includegraphics*[width=8.6cm]{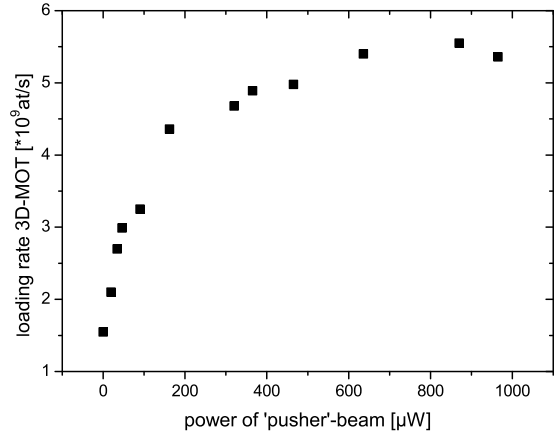}

\caption{3D-MOT loading rate depending on the power of the additional ''pusher''-beam for the 2D-MOT. An enhancement of the flux and the loading rate at
the same time is possible.}

\label{fig:5}

\end{figure}

With the normal parameters used in everyday operation (total laser power 2D-MOT $130\,$mW, total laser power 3D-MOT $60\,$mW, Rb-reservoir temperature 90$^{\circ}$C) and the
additional ''pusher''-beam, we obtain a 3D-MOT loading rate
of $5.6\times 10^{9}\,$at/s, as shown in Fig.~\ref{fig:5}. With these parameters, the maximum 3D-MOT atom number we obtain is $2\times 10^{9}$ atoms, which saturates due to collisions with background (non-Rb) atoms. Operating the source with all important parameters deeper in the saturated regime, e.g. laser power or Rb vapor pressure, we can obtain loading rates $>10^{10}\,$at/s.

\section{Atomic launch analysis} \label{sec:atomiclaunch}
In this section, we investigate the features of the emitted atomic ensemble relevant for the determination of the scaling factor such as the accuracy of the forward velocity and pointing and orientation of the source, as well as the parameters determining the effective signal of the interferometer, i.e. the atomic temperature.

\begin{figure}[t]

\centering

\includegraphics*[width=8.6cm]{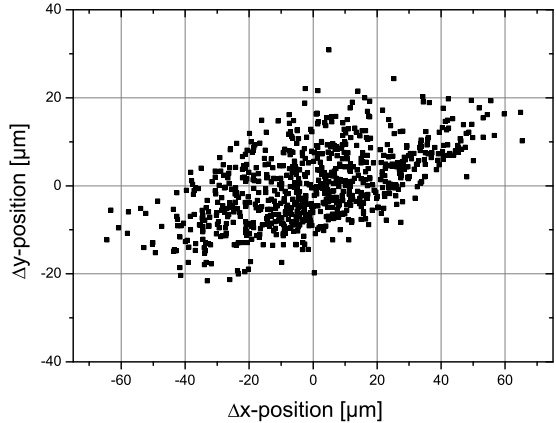}

\caption{Position of the atomic cloud in the MOT just before launching. The position is displayed for each dimension
relative to the mean value.}

\label{fig:6}

\end{figure}

To characterize the stability of the starting conditions for the moving molasses, we record the spatial position of the atoms just before the launch. For this, in situ images of
the fluorescence of the atoms trapped in the MOT are imaged onto a CCD camera over 30 minutes while cycling the experiment about 800 times. The result is displayed in Fig.~\ref{fig:6}. The recorded camera images provided information about the clouds position along the atomic beam propagation axis (''x''-dimension) as well as in the direction along the interferometers Raman beam-splitter propagation (''y''-dimension). We find standard deviations of $\sigma_x=23\,\mu$m and $\sigma_y=8.5\,\mu$m for the variations of the spatial position in the two dimensions mentioned before. To convert the measured position variations into an uncertainty for the resolution of rotations of the interferometer, the aberrations of the interferometer beam-splitters have to be taken into account \cite{Ref14}. Considering the aberrations given in Ref.~\cite{Ref14}, we can estimate the position variations to induce an uncertainty in the rotation measurement on the order of less than few nrad/s, assuming our projected enclosed area A of $22\,$mm$^2$.

\begin{figure}[t]

\centering

\includegraphics*[width=8.6cm]{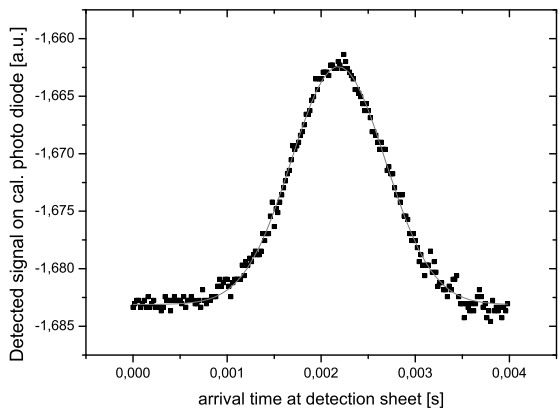}

\caption{The atomic velocity and temperature are measured in the interferometer zone in a time-of-flight measurement.
The arrival time at a thin sheet of light is recorded and can be converted to a spatial distribution because of the
known atomic mean velocity.}

\label{fig:7}

\end{figure}

\begin{figure}[t]

\centering

\includegraphics*[width=8.6cm]{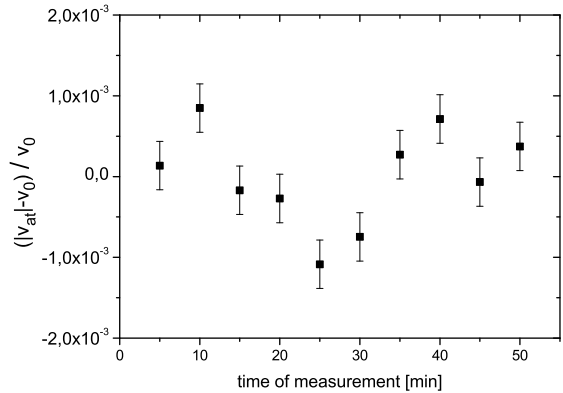}

\caption{Relative stability of the measured atomic velocity for cycling time of $50\,$minutes. The actual velocity $|v_{at}|$ is compared relative to the value $v_0$, which is calculated from the applied frequency difference for the moving molasses. Each displayed point is the mean value of consecutive measurements, see text for details.}

\label{fig:8}

\end{figure}

In the next step of the launch analysis we examine the stability of the starting velocity $\vec{v}_{at}$, beginning with measuring the absolute value $|\vec{v}_{at}|$. The variation of $|\vec{v}_{at}|$ has been measured with a time-of-flight method. For this measurement the arrival times of the atomic wave packets was measured at various positions downstream in the interferometer chamber. Each detection zone consists of a thin
($0.5\,$mm) sheet of retro-reflected $\sigma^+$-polarized light, resonant with the $|F=2>$$\rightarrow$$|F'=3>$) transition
of the $D_2$-line. The time dependent
fluorescence resulting when launching the atoms through the detection zones is recorded by a calibrated photo diode at each zone. A typical measurement curve is shown in Fig.~\ref{fig:7}. We measure
the spatial distance between the detection zones and calculate the absolute velocity $|\vec{v}_{at}|$. This measurement setup is not sensitive to the direction of $\vec{v}_{at}$ (for small misalignments from the perfect direction), because the
detection light is perpendicular to the ideal horizontal trajectory. Within the uncertainty of the distance measurement, we find no significant difference between the theoretically calculated and the measured velocity. To analyze the stability, consecutive
velocity measurements are performed during a 50 minutes experimental cycling. Each measurement point consists of the
mean velocity value of 5 min averaging. The result is displayed in Fig.~\ref{fig:8}: The relative standard deviation of the velocity
$\sigma_{|\vec{v}_{at}|}$ is $3\cdot 10^{-4}$ (which is also the uncertainty of the velocity measurement), with the biggest variation being $1,1\cdot10^{-3}$. Additionally, the measured absolute velocity value $|\vec{v}_{at}|$ is the same for both sources to a relative uncertainty of $<3\times10^{-4}$. Assuming the projected enclosed area of our future gyroscope of $22\,$mm$^2$ the measured variations of the absolute velocity would lead to a rotation uncertainty of $5\times 10^{-9}\,$rad/s. At the same time, the variations of the absolute velocity will lead to changes in the spatial position of the atomic cloud of about $50\,\mu$m at the beam splitter position. Analogous to the variation of the clouds starting position, we estimate this effect to induce a rotation uncertainty of few nrad/s.

The same experimental setup is also used to determine the temperature of the atomic cloud: Thanks to the precisely measured mean velocity, we can convert the time-dependent photo diode signal into a spatial dependence and deduce the diameter of the atomic cloud at different times after the launch. We can neglect the finite size of the light sheet (which is small compared to the $e^{-1/2}$-cloud size of $4.5\,$mm) and assume an isotropic distribution. From this time-of-flight (TOF) experiment we conclude the atomic
temperature with optimized launching parameters to be 8$\,\mu$K. The optimization of the temperature is based on an
adiabatic release of the atoms \cite{Ref9}. We find as optimal release procedure a ramp down of the cooling laser power at the end of the launch sequence
in about $0.8\,$ms with a detuning of the cooling lasers of about $-12\,\Gamma$ (in addition to the differential detuning of each ''moving'' beam).
\begin{figure}[t]

\centering

\includegraphics*[width=6.6cm]{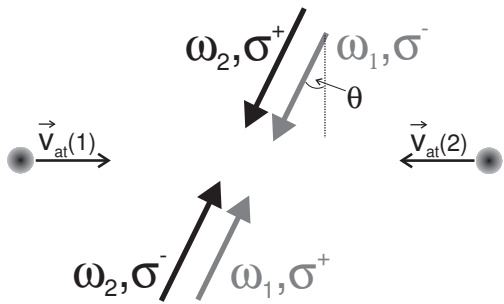}

\caption{Schematic of the experimental setup for velocity sensitive Raman-transitions, top-view. The two atomic ensembles launched with $\vec{v}_{at}$(1,2) interact from two directions with each of the two Raman laser beams with frequency $\omega_{1,2}$. A transition is only possible with equal polarization in the two Raman laser beams ($\sigma^+-\sigma^+$ or $\sigma^--\sigma^-$) which is used to suppress Doppler insensitive transitions thanks to the chosen polarization configuration. Each of the two atomic ensembles can undergo a resonant transition either for negative or positive detuning relative to the unperturbed atomic transition with one of the resonant beam pairs due to the Doppler effect.}

\label{fig:11}

\end{figure}

\begin{figure}[t]

\centering

\includegraphics*[width=8.6cm]{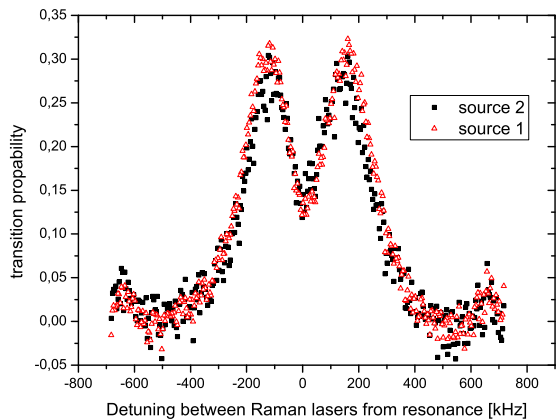}

\caption{Variation of the frequency difference between the Raman lasers for a velocity selective pulse, compared to the
unshifted ''clock''-transition. The pulse length of about $10\,\mu s$ was chosen to give maximal transition amplitude for optimal detuning.}

\label{fig:9}

\end{figure}

\begin{figure}[t]

\centering

\includegraphics*[width=8.6cm]{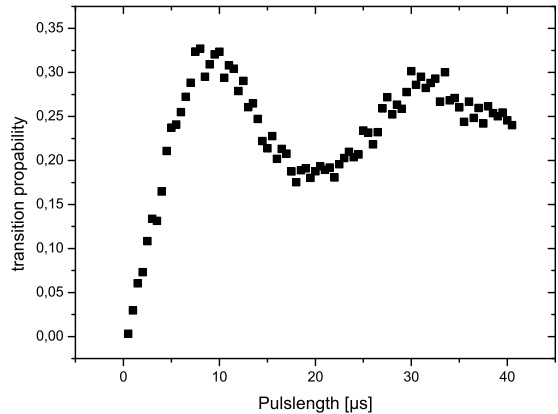}

\caption{Variation of the pulse length of a velocity selective Raman transition for optimal detuning (inferred from
Fig.~\ref{fig:9}). The maximal transition probability is related to the atomic temperature, which can be determined by this
measurement, in this case 10$\,\mu$K.}

\label{fig:10}

\end{figure}

The atomic temperature is also determined in a second complementary measurement by deducing the temperature dependent efficiency of Doppler-sensitive two-photon Raman-transitions. For this, a preceding internal atomic state preparation into $|F=1,m_F=0>$ is applied following the launch. We perform the velocity selective Raman-transitions between the $|F=1,m_F=0>$ and the $|F=2,m_F=0>$ sub-states of the $5\:^{2}S_{1/2}$ level, the so-called ''clock-transition'' at about $6,834\,$GHz. We use two phase-stable, perfectly overlapped laser fields of frequency $\omega_1$ and $\omega_2$, with a
frequency difference of about $6,8\,$GHz to drive the Raman transition. To be Doppler sensitive, both beams interact with the atoms from two directions, see Fig.~\ref{fig:11}. The laser polarizations are chosen such, that only counter-propagating beam pairs can drive a Raman transition. The laser beams, precisely leveled perpendicular to gravity, cross the atomic trajectories nearly perpendicular, but with a small angle deviation $\theta$ (with $\vartheta=90^\circ -\theta$), allowing a selection of one of the two driving pairs due to the Doppler effect. The normalized excitation spectrum is obtained by detecting consecutively both internal states. The interaction of the two counter-propagating wave packets and the two Raman-beam pairs results in a double resonance as shown in Fig.~\ref{fig:9}. To determine the maximal obtainable transition probability, we perform Rabi oscillations between the two atomic states, by tuning the Raman-lasers on one of the two resonances and varying the pulse length, see Fig.~\ref{fig:10}. We obtain a maximal transition amplitude of 0.33. This result is compared to a theoretical simulation of the temperature-dependent reduction of the transition probability including the dependence to the experimental parameters (like the driving field Rabi frequency). For this, the reduction of the transition amplitude due to other effects than the temperature is excluded with an independent Doppler-insensitive measurement. In this way, we find a temperature of 10$\,\mu$K, slightly higher than in the TOF
measurement. We address the state preparation for the interferometer, which includes optical pumping, to be responsible
for a small additional heating, explaining the difference between the two independent measurements.

The positions of the frequency dependent maxima of the Raman transition shown in Fig.~\ref{fig:9} provides additional information about the overlap of the atomic trajectories of the both source units. In addition to the detuning of about $130\,$kHz, which results from the intentionally applied angle $\theta\simeq$12mrad, the maxima for the two sources are displaced by about $15\,$kHz, which accords to an angle difference of the atomic trajectories in the horizontal plane of about $1.34\,$mrad. We can assign the frequency difference of both maxima to a pure angle displacement, because both absolute horizontal velocities $|\vec{v}_{at}|$ have been measured to be equal to much better precision. For the determined angle deviation, we mainly make an undesired, independently determined tilt of the experimental platform during the measurement responsible. Due to the tilt of about $5.2\,$mrad (with respect to the horizon) the small, ideally vertical atomic velocity component we apply to realize parabolic atomic trajectories is partly converted into a horizontal velocity component. The tilt of the platform will be better controlled for future experiments, either due to passive stability or an active servo loop. Thus subtracting the tilt effect, the two clouds trajectories are parallel to $<0.2\,$mrad, with the residual angle deviation resulting from molasses laser beam misalignment. Converting the atomic velocity direction misalignment into changes in the area A (for the Sagnac measurement) results in a relative variation of only $10^{-6}$ for A assuming $\vartheta\approx 90^{\circ}$, changing the measured rotation rate in our interferometer by much less than nrad/s.

The pointing misalignment in the vertical direction was not measured independently, but we estimate the misalignment in the vertical direction to be of the same order as in the horizontal direction, assuming the same accuracy of alignment for all molasses laser beams. The position variation of the atomic cloud at the Raman beam-splitters induced by this estimated horizontal misalignment is about 40$\,\mu$m, corresponding to a rotation error of few nrad/s, again following the results from Ref.~\cite{Ref7}.

\section{Conclusion and Outlook}
\label{sec:conclusion} 
In this article we have presented a compact source system comprising a 2D-MOT replacing a Zeeman-slower and a 3D-MOT/moving molasses providing precise and stable launch conditions as required for atomic Sagnac interferometers. The flux of the 2D-MOT is comparable to thermal sources and the 3D-MOT loading flux saturates above $10^{10}\,$at/s.
Under usual operating conditions we load the 3D-MOT with $10^9$ atoms in 200ms. With our measured state preparation efficiency of 70$\%$ and the measured efficiency of the Doppler-sensitive Raman transitions of 33$\%$ we will be able to operate the interferometer at a cycling rate of 3 Hz with $3\times 10^8\,$at/s contributing to the measurement signal.

Aiming for resolutions of rotations of better than $10^{-8}\,$rad/s, the performed measurements show that the performance of the realized atomic source will not be a limitation for the short-term sensitivity of the interferometer. However, a further increase of the resolution of rotations or the enhancement of the long-term stability require an improvement of the relevant parameters like the velocity stability or the pointing of the source. In this respect, the control of the corresponding molasses parameters, like the laser power or laser beam alignment, is currently improved. This measures should additionally allow a further reduction of the atomic temperature. Moreover, our source design permits the planned extension of the velocity control for the two sources, allowing a tuning of both relevant velocity components for each source individually, which lead to a further optimization of the overlap of the atomic trajectories.

Beside the described application for interferometry, the presented source concept could also be adapted for other kinds of atom optic experiments, where a brilliant atomic source with flexible and precisely controlled atomic velocity in the range of several m$/$s is desired, for example the injection of atoms into a guiding potential \cite{Ref19}.

Thanks to the compactness of our setup we will be able to realize the final atomic sensor as a transportable device. This will be an interesting option for combined measurements with other high
resolution atomic or optical sensors \cite{Ref18}, allowing interesting studies in applied sciences like geology or geodesy such as the search for variations of the earth rotation. 

\section{Acknowledgments}
This work is supported as part of the SFB 407 of the ''Deutsche Forschungsgemeinschaft'' and as part of the FINAQS cooperation of the European union.

\end{document}